# Dynamic Modeling for Representing Access Control Policies Effect

Kambiz Ghazinour
Department of Computer Science
Kent State University
Kent, Ohio, USA
kghazino@kent.edu

Mehdi Ghayoumi
Department of Computer Science
Kent State University
Kent, Ohio, USA
mghayoum@kent.edu

## Abstract

*In large databases, creating user interfaces for browsing or performing insertion, deletion or modification of data is very costly in terms of programming. In addition, each modification of an access control policy causes many potential and unpredictable side effects which cause rule conflicts or security breaches that affect the corresponding user interfaces as well. While changes to access control policies in databases are inevitable, having a dynamic system that generates interfaces according to the latest access control policies becomes increasingly valuable. Lack of such a system leads to unauthorized access to data and eventually violates the privacy of data owners. In this work, we discuss a dynamic interface that applies Role Based Access Control (RBAC) policies as the output of policy analysis and limits the amount of information that users have access to according to the policies defined for roles. This interface also shows security administrators the effect of their changes from the user's point of view while minimizing the cost by generating the interface automatically.*

## 1. Introduction

Enforcing access control is a crucial issue in all computer systems. Using access control mechanisms guarantees that malicious and questionable users cannot access sensitive data and also legitimate users cannot accidentally access parts of the data that are not supposed to be revealed. In large databases, where we may have hundreds of different roles and access control policies, handling RBAC policies is even harder [4]. Furthermore, programming user interfaces that conform to the latest dynamic access control policies is not generally a straightforward job for the following reasons: first, fields may be added to the tables in the database after the interface has been designed. Hence, the user interface must be redesigned again to represent the data included in newly added fields. Second, as the number of users and roles increase in the database, it becomes difficult to program different user interfaces for each role or user.

Although applying RBAC [7] facilitates managing access control policies more efficiently than conventional access control methods such as Mandatory Access Control (MAC) and Discretionary Access Control (DAC) [6], designing a dynamic interface that conforms to the access control policies needs more work.

### 1.1. Contribution of this paper

In this work, we introduce a model that creates forms dynamically based on the tables' structures and the access policies in the relational database management system (RDBMS). This approach reduces the extensive amount of work needed to rebuild user interfaces based on the access control policies statically. Furthermore, this approach enables the security officers and designers to have the opportunity for immediate testing to see if roles are working as they should. This contribution is discussed in Sections 3 and 4 in which we introduce our approach and the dynamic user interface developed based on it. To illustrate the functionality of the dynamic interface, Section 5 describes an example to demonstrate our application. Section 6 concludes the paper and give possible future research directions to extend this idea.

## 2. Background and Related Work

In RBAC [7], object accesses are controlled by roles (or job functions) in an enterprise rather than a user or a group. RBAC, as an alternative to conventional DAC and MAC mechanisms, is required for handling data authorization management in a complex environment as has been discussed in the literature [7]. RBAC has been introduced as a cost effective access control mechanism [6]. Due to its characteristics (i.e. rich specification, separation of duty and ease of management), it is being employed in a large variety of domains [8].

In RBAC, the main goal is to provide a model and tool to help manage access controls in an enterprise with a very large number of users and data items. The main components of RBAC are roles, users and permissions where role represents job functions, and permissions are defined on objects and operations. In particular, permissions can be defined in terms of allowing or preventing a role from performing a specific action on a specific data object.

There have been many extensions of RBAC introduced in the literature. For instance, Byun and Li [2] introduced a purpose-based access control for privacy protection in relational database systems which is based on Role-based access cotrol model. As another example, Dafa-Alla et al., [3] introduced PRBAC: An Extended Role Based Access Control for Privacy Preserving Data Mining.

Although, according to the National Institute of Standard and Technology (NIST) standard, there are different levels of RBAC including flat, hierarchy, constrained, and symmetric options [8], in this work, we focus on the flat model and leave the application of other techniques as they are really extensions for future work.

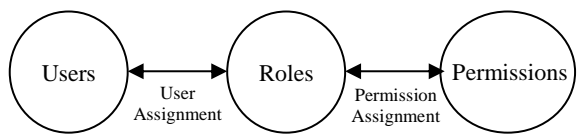

**Figure 1. Flat RBAC model**

In flat RBAC, permissions are assigned to each role and users are assigned to one or more roles as shown in Figure 1. A user is granted access to an object when the user is active in a role that has the required permissions. For instance, these two tuples

(Staff, +, Read, email)
(Staff, -, Update, name)

mean any user with the Staff role has the privilege to read the field *email* but not to update the field *name*. Some of the access control policies can result in a possible access rules conflict that affects the access level of the user. Vaniea *et al.* [9] discuss an interface that visualizes the output of policy analysis and helps security professionals find conflicting policies. In our work, we introduce a software package that dynamically creates user interfaces based on the user's latest access control privileges. We believe this dynamic user interface reduces the extensive amount of work needed to rebuild user interfaces based on the access control policies statically.

Agrawal *et al.* [1] propose a language construct and implementation design that restricts the queries submitted to the RDBMS to enforce privacy policies.

In their solution, fields that the user requests to see but does not have privilege to access, are returned with a null value. This is different from our approach in which the user only observes fields for which they have privilege to see. Therefore, information about existence of the field(s) is not revealed to malicious users.

## 3. Our Approach

In this section, we describe our approach to creating a dynamic user interface (DUI) based on user access control. As illustrated in Figure 2, our model consists of two main engines, Component Manager and RBAC Extractor. The data flow is described as follows.

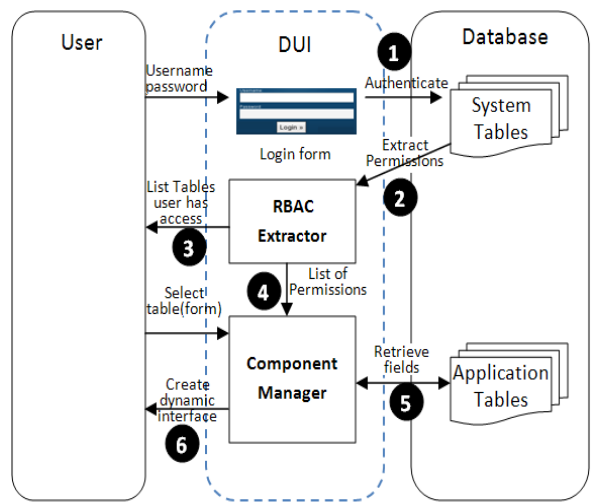

**Figure 2. Dynamic interface data flow and architecture**

1. When the user wants to log in to our software, the user name and password should match the ones entered into the database by the security manager or database administrator.

2. After the user authentication is complete, by reviewing the User Assignment relationship, the RBAC Extractor engine determines roles that are assigned to the user by the Security Administrator. According to the RBAC architecture [7], each user is associated with at least one role.

3. The list of tables the user is allowed to observe are then displayed and they specify the form they will use. This form is related to one or more table(s) in the database.

4. The RBAC Extractor engine reviews the Permission Assignment relationship to identify what permissions are assigned to the related roles of the user. After considering all the permissions, RBAC Extractor provides the Component Manager with a list of permissions for four different actions, *Select*, *Update*, *Insert* and *Delete*. These could be extended to other operations but we limit our discussion to these four operations to illustrate our system.

5. The Component Manager identifies the related fields on the desirable table(s) and their specifications.

6. Finally, the dynamic user interface is created and displayed to the user based on the data collected through steps 4 and 5. If the user selects another table from the list, data flow starts from step 4.

To clarify the task of the Component Manager, consider the following example. The user wants to work with a form related to table *T* which has fields *a*, *b*, *c* and *d*. The RBAC Extractor engine determines that according to the access control policies this user cannot see (or select) fields *b* and *c* from this table since accessing them is prohibited by one of the user's roles. Hence, the Component Manager only shows fields *a* and *d* on the form to the user. This approach is also known as query rewriting and has attracted much attention in the literature [1, 5]. The same procedure is followed for other actions such as *Delete*, *Update* and *Insert* which affects the user interface as well. We will discuss each of them in detail in Section 4.

## 4. Dynamic User Interface

In this section, we describe our system and demonstrate its utility. We focused on Microsoft SQL Server 2000 and 2005 as a well known RDBMS that supports RBAC mechanism. We also used Microsoft Visual Basic .NET 2003 to develop the system. Through ADO .NET the program connects to the RDBMS and extracts information about users, roles and permissions stored in system tables.

Generally, in a RDBMS such as MS SQL Server, to enforce access control policies the following steps are required of the Security Administrator. First, users and roles (job functions) are defined. Second, for each role, privileges to access different objects (tables and related fields) are defined. Third, one or more roles are assigned to each user. There are two forms of permission, grant and deny. (Deny is different than revoking a permission, and it essentially means a negative permission). If the user has a role, and within that role they are granted access (i.e. *Select*) to a specific field or table, then the user's (*Select*) query on the table returns proper results. On the other hand, if the user is not privileged to perform the specific action on that field or is denied access to that field, then the result of the query will be an error message that indicates insufficient permissions to *Select* this field. The same rules apply to *Insert, Delete,* and *Update* permissions.

Handling permissions of one role for a user seems straightforward since no conflict can occur. However, often the user has more than one role, and those roles may contradict one another. For instance, imagine that Alice has two roles called *Role1* and *Role2*. According to *Role1,* she has access to the field *CustomerID* from the *Customer* table and according to *Role2,* she is not allowed to see this field. According to the security policies defined in MS SQL Server system table, Alice is not allowed to see that field because the deny permission dominates the grant permission. There are other possible ways to create contradicting permissions such as granting access to the whole table for one role and denying access to specific field(s) of the same table for another role where a user has both of the roles. The algorithm that determines the resulting combination of the permissions the user has due to their corresponding roles $R_1$ to $R_n$ on a specific field has the following pattern:

ResultingPermission($R_1,…,R_n$)
1. result = deny;
2. For all the roles from $R_1$ to $R_n$
    If there exist a deny permission then
        result = deny and exit
    Else if there exist a grant permission then
        result = grant
3. Return result

Thus, this system is implemented to enforce the access control policies and different combinations of the roles defined in RBAC systems. When the user logs in, the RBAC Extractor engine identifies the user from the list of available users in the RDBMS. All the roles associated with the user are then extracted from the system tables. Using the above algorithm, the engine then determines conflicting parts and generates a list of permissions that covers all the roles the user has and provides the Component Manager engine with that list. To clarify this, consider a user who has two roles called *Role1* and *Role2*. The policies for permissions to access tables and fields are shown in Table 1. The symbol ✕ represents deny and ✓ represents grant access. If access to a field is not defined by any role (eg. the *Address* field in *Customers* table) then the user should not have access to that field. In other words, the user does not have access to a field unless explicitly granted the privilege.

In our software, we have four classes, CDB, CTable, CGrid, and CSet. Class CDB is responsible for connecting to the RDBMS and submitting queries to the database. CTable is a subdivision of CDB which obtains the name of a specific table in the database and extracts information about the fields of that table. It also connects to the metadata and extracts the information about RBAC policies. In other words, having the name of a specific table and the user, this class determines what permissions are assigned to the user to access that specific table. CGrid and CSet are two classes that deal with the dynamic user interface itself. CGrid is used to display records of a table in the form of a grid. CSet is in charge of managing the components of a form related to one or more tables.

**Table 1. Sample of conflict resolution domination**

| Table | Field | R1 | R2 | Result |
|---|---|---|---|---|
| Customers | CustomerID | ✓ | ✓ | ✓ |
| Customers | CustomerName |  | ✓ | ✓ |
| Customers | Address |  |  | ✗ |
| Employees | EmployeeID | ✗ | ✓ | ✗ |
| Employees | EmployeeName | ✓ |  | ✓ |
| Employees | Phone |  | ✗ | ✗ |
| Orders | EID | ✗ | ✗ | ✗ |
| Orders | CID | ✗ | ✗ | ✗ |
| Orders | OrderDate | ✓ | ✓ | ✓ |
| Orders | Payment | ✓ | ✓ | ✓ |

To clarify the use of CGrid and CSet we define that CGrid is used for the *Select, Update and Delete* actions whereas the CSet is used when we want to *Insert* a new record in a table and we need a form that contains all the corresponding fields.

The above classes interact with the dynamic user interface catalogues that we add to the RDBMS. Although there are some features of the fields such as *allow null* and *data type* that can be recognized from the tables themselves, there are other features that need to be controlled by the user interface as well. Some of these features are location of the components on the form, enforcing the accurate data type entry by the user, component's visibility on the form, and so on.

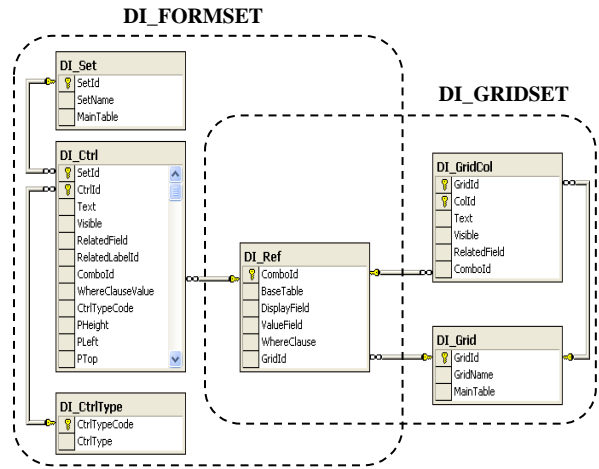

**Figure 3. Dynamic interface catalogues**

Generally, these features are hard coded in an executable program which makes a static user interface. To represent the semantics of our model we propose user interface catalogues in which we store component features in the database to be retrieved on the user's demand. The software then uses these parameters to create the proper user interface. These catalogues, DI_FORMSET and DI_GRIDSET, represent information about rows and columns in a grid and components in a tabular form. We have implemented these catalogues as a set of tables added to the RDBMS. Figure 3 illustrates these catalogues and the corresponding tables. DI_FORMSET contains DI_Set, DI_Ctrl, DI_CtrlType, and DI_Ref tables. DI_GRIDSET, in addition to tables DI_Grid and DI_GridCol, shares DI_Ref with DI_GRIDSET.

When a form is selected by the user, the system refers to the tables DI_Set and DI_Ctrl and extracts all the information related to that specific table. In the next step, all the components of the form are located and their properties are set according to the information derived from DI_Set and DI_Ctrl. It is clear that in this phase, tables DI_CtrlType and DI_REF help the system to enforce the correct data types of the components and referring tables, respectively. When a form is required to be in the form of a grid, DI_GRIDSET catalogue and its corresponding tables provide information to the CGrid class to illustrate data in a grid. It should be mentioned that to fill in the above tables we have prepared a simple user interface called DI_Creator. Using this software, the Security Administrator can define initial features of each element on the form according to the specifications described in the design phase of the application development. Section 5 will present an example to clarify the task of each table and class.

## 5. Example

Alice has both Staff and Advisor roles in a company. Based on her roles and the access control policies defined by the Security Administrator of the company, she has corresponding accesses of *Select*, *Insert*, *Update* and *Delete* on different fields and tables. For instance, as shown in Figure 4 (a) and (b), role Advisor can *Select* all of the fields in the *Customer* table. However, role Staff can *Select* the field *City*, and is not allowed to see other columns of the customer table (except the field *CompanyName* where no grant or deny permission is explicitly specified).

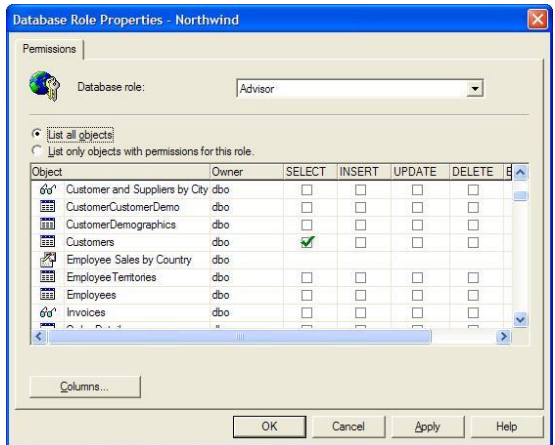

**(a) Advisor's access control policy**

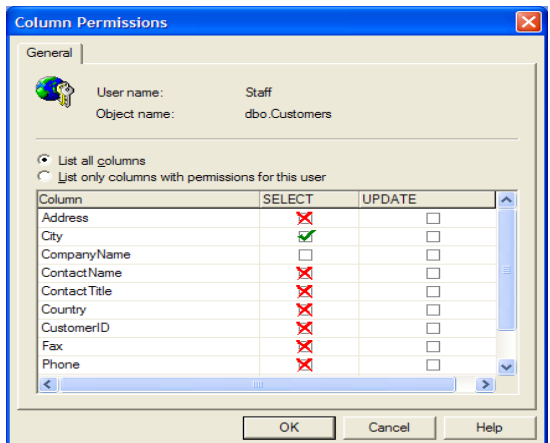

**(b) Staff's access control policy**

**Figure 4. Access control for roles Advisor and Staff**

Since Alice has both Staff and Advisor roles, and we assume that negated permission is dominant, according to the combinations discussed in Table 1, she is only able to see the fields *City* and *CompanyName*. As described in Section 3, the RBAC

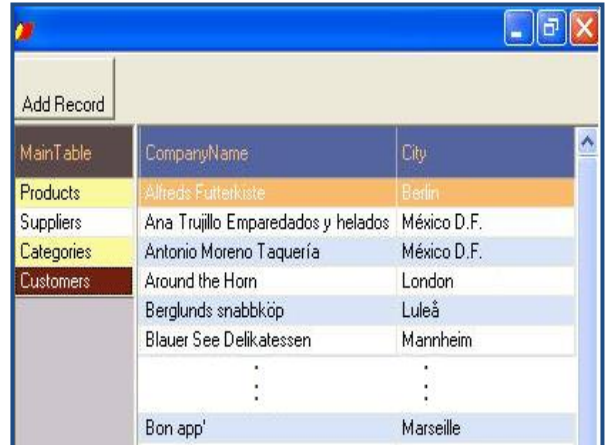

**Figure 5. Dynamic user interface created for Alice**

Extractor engine is responsible for finding the right combination of roles and provides the results to the Component Manager. In this example, the Component Manager needs to show the above two fields in the *Customers* table to Alice. Hence, when she logs in to the system and clicks on the *Customers* form she sees the user interface shown in Figure 5.

## 6. Conclusion and Future Work

We have presented a model for generating dynamic interfaces based on the RBAC policy. This approach can be extremely useful in large databases used in enterprises where a large amount of resources are spent to design, develop and maintain the user interfaces. Since this model dynamically creates the user interfaces and also enforces the latest RBAC policies, it saves a considerable amount of time and cost when producing middleware that works with databases. From the privacy point of view, unlike current approaches, our work does not reveal the existence of fields to the users who are not privileged to access.

Our future work is to extend the techniques presented in this paper to create a dynamic user interface for applying privacy policies stored in privacy preserving database systems.

In future work, developers can add their own class to the software to support more features. In this case, options like reporting and printing can be added to the system. Also, it would be an interesting project to use this methodology in web-based applications as well.

Another interesting future research direction that we are working on is to incorporate the notion of trust in to the model. For example, when a user has several unnecessary accesses to a piece of data, the systems learns this behavior and reduces the level of trust in that particular user. The lack of trust results in automatic modification of the policies in place which reduces the privilege(s) given to the user. Once the policy is modified, it effects the user interface as well. In brief, unnecessary and redundant access to data items result in losing the privilege of accessing them.

## 7. References

1. R. Agrawal, P. M. Bird, T.W.A. Grandison, G. G. Kiernan, S. I. Logan, & W. Rjaibi. *Extending Relational Database Systems to Automatically Enforce Privacy* Policies, 2007.

2. J. Byun, and N. Li. "Purpose based access control for privacy protection in relational database systems." The VLDB Journal 17.4 2008: 603-619.

3. A. F.A. Dafa-Alla, E. H. Kim, K. H. Ryu, Y. J. Heo. "PRBAC: An Extended Role Based Access Control for Privacy Preserving Data Mining" In Proceedings of the Fourth Annual ACIS International Conference on Computer and Information Science (ICIS'05) of IEEE, 2005.

4. D. Ferraiolo, D. Kuhn, and R. Chandramouli. Role-based access control Artech House computer security series. Artech House, 2003.

5. S. Mohan, A. Sengupta, & Y. Wu. Access control for XML: A dynamic query rewriting approach. *Proceedings of the 14th ACM International Conference on Information and Knowledge Management,* 251-252., 2005.

6. R. Ramakrishnan and J. Gehrke. Database Management Systems. McGraw-Hill Science / Engineering / Math, 2003.

7. R. Sandhu, D. Ferraiolo, and R. Kuhn. The NIST model for role-based access control: towards a unified standard. Symposium on Access Control Models and Technologies: Proceedings of the fifth ACM workshop on Role-based access control, 26(28):47–63, 2000.

8. The Economic Impact of Role-Based Access Control, RTI Project Number: 07007.012, National Institute of Standards and Technology (NIST), 2002. [Online]. Available:http://www.nist.gov/director/prog-ofc/report02-1.pdf

9. K. Vaniea, Q. Ni, L. Cranor & E. Bertino. Access control policy analysis and visualization tools for security professionals. SOUPS Workshop (USM) 2008.